\documentclass{ieeeaccess}
\usepackage{cite}
\usepackage{amsmath,amssymb,amsfonts}
\usepackage{algorithmic}
\usepackage{graphicx}%
\usepackage{url}

\usepackage[colorlinks=true,bookmarks=false,citecolor=blue,urlcolor=blue]{hyperref} %
\usepackage{thmtools}

\usepackage{textcomp}
\usepackage{bm}
\usepackage{multirow}
\usepackage{float}
\usepackage{braket}
\usepackage{makecell}
\usepackage{booktabs}
\usepackage{ifthen}
\usepackage{mathtools}
\usepackage{acronym}

\renewcommand{\vec}[1]{\bm{#1}}

\newcommand{\me}{\mathrm{e}}

\usepackage{pifont}%

\usepackage{bm}
\renewcommand{\vec}[1]{\bm{#1}}
\usepackage{bbm}

\newcommand{\trans}[1]{#1^{\mathsf{T}}}

\newcommand{\lto}{\leftarrow}

\newcommand\blfootnote[1]{%
	\begingroup
	\renewcommand\thefootnote{}\footnote{#1}%
	\addtocounter{footnote}{-1}%
	\endgroup
}

\definecolor{amber}{RGB}{227, 150, 16}

\newcommand{\REV}[1]{#1}

\makeatletter
\AtBeginDocument{\DeclareMathVersion{bold}
	\SetSymbolFont{operators}{bold}{T1}{times}{b}{n}
	\SetSymbolFont{NewLetters}{bold}{T1}{times}{b}{it}
	\SetMathAlphabet{\mathrm}{bold}{T1}{times}{b}{n}
	\SetMathAlphabet{\mathit}{bold}{T1}{times}{b}{it}
	\SetMathAlphabet{\mathbf}{bold}{T1}{times}{b}{n}
	\SetMathAlphabet{\mathtt}{bold}{OT1}{pcr}{b}{n}
	\SetSymbolFont{symbols}{bold}{OMS}{cmsy}{b}{n}
	\renewcommand\boldmath{\@nomath\boldmath\mathversion{bold}}}
\makeatother

\def\BibTeX{{\rm B\kern-.05em{\sc i\kern-.025em b}\kern-.08em
		T\kern-.1667em\lower.7ex\hbox{E}\kern-.125emX}}

\begin{document}
	
	\acrodef{BCH}{Bose--Chaudhuri--Hocquenghem}
	\acrodef{FER}{frame error rate}
	\acrodef{BSC}{binary symmetric channel}
	\acrodef{DE}{density evolution}
	\acrodef{LDPC}{low-density parity-check}
	\acrodef{QLDPC}{quantum low-density parity-check}
	\acrodef{VN}{variable node}
	\acrodef{CN}{check node}
	\acrodef{BP}{belief propagation}
	\acrodef{BP2}{binary belief propagation}
	\acrodef{BP4}{quaternary belief propagation}
	\acrodef{NBP}{neural \ac{BP}}
	\acrodef{NBP4}{neural \ac{BP4}}
	\acrodef{NBP2}{neural \ac{BP2}}
	\acrodef{OBP4}{\ac{BP4} using overcomplete check matrix}
	\acrodef{NOBP4}{\ac{NBP4} using an overcomplete check matrix}
	\acrodef{NBP}{neural belief propagation}
	\acrodef{GB}{generalized bicycle}
	\acrodef{PCM}{parity check matrix}
	\acrodefplural{PCM}[PCMs]{parity check matrices}
	\acrodef{OSD}{ordered statistics decoding}
	\acrodef{QEC}{quantum error correction}
	\acrodef{QSC}{quantum stabilizer code}
	\acrodef{CSS}{Calderbank–Shor–Steane}
	\acrodef{NN}{neural network}
	\acrodef{LLR}{log-likelihood ratio}
	\acrodef{MWPM}{minimum weight perfect matching}
	\acrodef{MS}{min-sum}

	\history{Date of publication 5 February 2025, date of current version 10 February 2025.}
	\doi{10.1109/ACCESS.2025.3539475
	}
	
	\title{Quaternary Neural Belief Propagation Decoding of Quantum LDPC Codes with Overcomplete Check Matrices}
	\author{\uppercase{Sisi Miao}\authorrefmark{1} \IEEEmembership{Student Member, IEEE},
		\uppercase{Alexander Schnerring}\authorrefmark{2}, \uppercase{Haizheng Li}\authorrefmark{1}, and \uppercase{Laurent Schmalen}\authorrefmark{1}
		\IEEEmembership{Fellow, IEEE}}
	
	\address[1]{Karlsruhe Institute of Technology (KIT), Communications Engineering Lab (CEL), 76187 Karlsruhe, Germany}
	\address[2]{\emph{now with} German Aerospace Center (DLR), Institute of Solar Research, 04005 Almería, Spain}
	\tfootnote{This work has received funding from the European Research Council (ERC) under the European Union's Horizon 2020 research and innovation programme (grant agreement No. 101001899).}
	
	\markboth
	{Sisi Miao \headeretal: Quaternary Neural Belief Propagation Decoding of Quantum LDPC Codes with Overcomplete Check Matrices}
	{Sisi Miao \headeretal: Quaternary Neural Belief Propagation Decoding of Quantum LDPC Codes with Overcomplete Check Matrices}
	
	\corresp{Corresponding author: Sisi Miao (e-mail: sisi.miao@kit.edu).}

	\begin{abstract}
		Quantum low-density parity-check (QLDPC) codes are promising candidates for error correction in quantum computers. One of the major challenges in implementing QLDPC codes in quantum computers is the lack of a universal decoder. In this work, we first propose to decode QLDPC codes with a belief propagation (BP) decoder operating on overcomplete check matrices. Then, we extend the neural BP (NBP) decoder, which was originally studied for suboptimal binary BP decoding of QLPDC codes, to quaternary BP decoders. Numerical simulation results demonstrate that both approaches as well as their combination yield a low-latency, high-performance decoder for several short to moderate length QLDPC codes.
	\end{abstract}
	
	
	\titlepgskip=-21pt
	
	\maketitle
	
	\section{Introduction}
	\blfootnote{Parts of this work have been presented at the Information Theory Workshop (ITW), 2023~\cite{miao2023}.}\Ac{QEC} is an essential part of fault-tolerant quantum computing. Thanks to the potential of providing fault-tolerance with constant overhead~\cite{Gottesman14fault} and the recent breakthroughs in designing asymptotically good \ac{QLDPC} codes~\cite{tillich2013quantum,panteleev2021quantum,panteleev2022asymptotically,breuckmann21balanced}, \ac{QLDPC} codes are among the most promising candidates for future \ac{QEC} schemes.  
	
	One of the major challenges in implementing \ac{QLDPC} codes for practical quantum computers is the lack of a universal decoder that provides good decoding performance across a wide range of \ac{QLDPC} codes. Two main families of potential decoders have been explored in the literature. The first family is the class of graph-theory-based decoders, such as the \ac{MWPM} decoder~\cite{dennis2002topological,edmonds1965paths}, which is commonly used for topological codes. However, a significant drawback of this decoder is its decoding latency. The second family is the class of message-passing decoders such as the \ac{BP} decoder or its low-complexity variant, the \ac{MS} decoder. The decoding latency of such decoders is linear w.r.t. the block length and the number of decoding iterations. However, their excellent decoding performance for classical \ac{LDPC} codes is based on the condition that the Tanner graph has no short cycles (of length 4). This condition cannot be fulfilled in the case of \ac{QLDPC} codes, where $4$-cycles are unavoidable by construction. Therefore, for \ac{QEC}, it is crucial to modify the \ac{BP} decoder such that short cycles can be tolerated.
	
	A variety of methods have been proposed to address this issue, which can be categorized into two categories. The first category contains approaches that modify the \ac{BP} decoder itself, for example, message normalization and offsets~\cite{kuo2020refined,lai2021log}, layered scheduling~\cite{panteleev2021degenerate,raveendran2021trapping}, and matrix augmentation\cite{rigby2019modified}. The second category focuses on post-processing the output of the \ac{BP} decoder, e.g., \ac{OSD}\cite{panteleev2021degenerate,roffe2020decoding}, random perturbation~\cite{poulin2008iterative}, enhanced feedback~\cite{wang2012enhancedfeedback}, and stabilizer inactivation~\cite{crest2022stabilizer}. However, most of these post-processing methods introduce additional decoding latency, which makes them less appealing for \ac{QEC} where decoding has to be performed with ultra-low latency.

	In~\cite{miao2023}, we have proposed and studied two approaches to enhance the \ac{BP4} decoder for \ac{QLDPC} codes~\cite{DM98,DF07}. The first approach is to perform \ac{BP4} decoding on an overcomplete check matrix constructed by adding redundant low-weight checks to the original full-rank check matrix. The second approach is based on neural networks. We extend the \ac{NBP} decoder, which was originally proposed in~\cite{liu2019neural} for the \ac{BP2} decoder, to the \ac{BP4} decoder. \REV{In this work, we extend the results of~\cite{miao2023} by presenting an improved decoding strategy and including a detailed methodology for parameter optimization. A more comprehensive derivation and explanation of the proposed method is provided. The proposed decoder is also evaluated on various exemplary codes. Numerical simulation results demonstrate that both approaches as well as their combination yield a low-latency high-performance decoder for a wide range of short to moderate-length QLDPC codes. For example, for the evaluated toric codes, the proposed decoder achieves up to an order of magnitude reduction in post-decoding error rates compared to the \ac{MWPM} decoder.}
	
	\REV{The paper's primary contribution is the novel \ac{NBP} decoder for \ac{QLDPC} codes, specifically addressing the degeneracy of quantum codes. Additionally, the tailored loss function for BP4 decoders enables efficient training of the decoder weights.}
	
	The remainder of this paper is structured as follows: In Sec.~\ref{sec:preliminaries}, we briefly review the notions of \ac{QEC} with \acp{QSC}. Section~\ref{sec:BP} reviews the refined \ac{BP4} decoder and investigates the initialization of the \ac{BP} decoder. In Sec.~\ref{sec:overcomplete}, we introduce decoding with overcomplete check matrices. The construction method of the redundant checks is given, as well as a heuristic explanation for the performance improvement. In Sec.~\ref{sec:NBP}, the \ac{NBP} decoder is introduced based on the refined \ac{BP4} decoder. In Sec.~\ref{sec:results}, the proposed decoder is evaluated with numerical simulations and compared with the state-of-the-art decoding algorithms in the literature. Section~\ref{sec:conclusion} concludes this paper.
	
	\textbf{Notation}: We use boldface letters to denote vectors and matrices, e.g., $\vec{a}$ and $\vec{A}$. The $i$-th component of vector $\vec{a}$ is denoted by $a_i$, and the element in the $i$-th row and $j$-th column of $\vec{A}$ is denoted by $A_{i,j}$. Let $\vec{A}_i$ be the $i$-th row of a matrix $\vec{A}$. $\vec{A}^{\mathsf{T}}$ denotes the matrix transpose. Pauli operators corresponding to quaternary vectors are denoted with the same letter in calligraphic form, such as $\vec{e}$ and $\mathcal{E}$.  For brevity, the maximum number of decoding iterations $L$ of a decoder is given as a superscript, e.g., BP4$^{32}$ for $L=32$.

	\section{Preliminaries}
	\label{sec:preliminaries}
	\subsection{Stabilizer Codes}
	
	\Acp{QSC}~\cite{gottesman1997stabilizer,Calderbank1998quantum} are the quantum analogons of classical linear codes. To define a \ac{QSC}, we first need to define the Pauli operators. For simplicity, we ignore the global phase and consider the $n$-qubit Pauli group $\mathcal{G}_n$ consisting of Pauli operators on $n$ qubits $\vec{\mathcal{P}} = \mathcal{P}_1\otimes\mathcal{P}_2 \otimes \cdots \otimes \mathcal{P}_n$ where $\mathcal{P}_i\in \{\vec{I},\vec{X},\vec{Y},\vec{Z}\}$ are Pauli operators on the $i$-th qubit. Without risk of confusion, the tensor product $\otimes$ and the identity operator $\vec{I}$ can be omitted. The weight $w$ of a Pauli operator $\vec{\mathcal{P}}$ is the number of non-identity components in the tensor product. To find a valid stabilizer code, it is crucial to find the stabilizer group $\mathcal{S}$, which is an Abelian subgroup of $\mathcal{G}_n$. This problem can be transferred into a classical coding problem using the mapping of Pauli errors to binary strings $\vec{\mathcal{P}}\mapsto \begin{pmatrix}x_1\;\cdots\;x_n\mid z_1\;\cdots\;z_n\end{pmatrix}=:\begin{pmatrix}\vec{p}_{X}\mid \vec{p}_{Z}\end{pmatrix}$ with $\mathcal{P}_i = \vec{X}^{x_i}\vec{Z}^{z_i}$.
	We can check that Pauli errors $\mathcal{A}$ and $\mathcal{B}$ in $\mathcal{G}_n$ commute if the symplectic product of their corresponding binary strings $(\vec{a}_X \mid \vec{a}_Z)$ and $(\vec{b}_X \mid \vec{b}_Z)$ is 0, i.e.,
	\begin{equation}
		\sum_{i=1}^{n} a_{X,i}\cdot b_{Z,i}  + \sum_{i=1}^{n} a_{Z,i}\cdot b_{X,i}= 0,
		\label{eq:commute_binary}
	\end{equation}
	where the addition is performed in the binary field (modulo~$2$). Therefore, a stabilizer code can be constructed from a $[2n,k]$ classical binary code given by its full rank \ac{PCM} $\vec{H} = \begin{pmatrix}
		\vec{H}_X \mid \vec{H}_Z\\
	\end{pmatrix}$ of size $m=2n-k$ by $2n$ fulfilling the symplectic criterion:
	\begin{equation}
		\vec{H}_X \trans{\vec{H}_Z} + \vec{H}_Z\trans{\vec{H}_X} =  \vec{0}.
		\label{eq:symplectic criterion}
	\end{equation}
	\ac{CSS} codes are a special kind of stabilizer codes where 
	\[\vec{H} =\left(\begin{array}{c}
		\vec{H}_X'    \\
		\vec{0}
	\end{array}
	\middle\vert
	\begin{array}{c}
		\vec{0} \\  \vec{H}_Z'
	\end{array}
	\right).\]
	In this case, \eqref{eq:symplectic criterion} holds if $\vec{H}_X' \vec{H}_Z^{'\mathsf{T}} = \vec{0}$ and the code construction can be simplified. \Ac{QLDPC} codes are defined as a family of stabilizer codes whose row and column weights are upper bounded by a relatively small constant independent of the block length, i.e., $\vec{H}$ is sparse. Most of the promising \ac{QLDPC} codes constructed so far are \ac{CSS} codes and all the \ac{QLDPC} codes considered in this paper are \ac{CSS} codes.
	
	To jointly decode the four types of Pauli errors, it is convenient to consider the quaternary form of $\vec{H}$, denoted as $\vec{S}\in\text{GF(4)}^{m\times n}$, where GF(4) consists of the elements $\{0,1,\omega,\bar{\omega}\}$. $\vec{H}$ can be converted to $\vec{S}$ using the mapping of binary vectors to vectors over GF(4) $\begin{pmatrix}x_1\;\cdots\;x_n\mid z_1\;\cdots\;z_n\end{pmatrix}\mapsto \begin{pmatrix}p_1\;\cdots\;p_n\end{pmatrix}=:\vec{p}$ with $p_i =x_i \omega  +  z_i \bar{\omega}$. Then, we can check that for $\mathcal{A}$ and $\mathcal{B}$, mapped to vectors $\vec{a}$ and $\vec{b}$ over GF(4), \eqref{eq:commute_binary} is equivalent to  $\langle \vec{a},\vec{b} \rangle = \sum_{i=1}^{n}\langle a_{i}, b_{i}\rangle = 0$, where $\langle\cdot,\cdot \rangle$ denotes the \emph{trace Hermitian inner product} over GF(4), which is defined as $\text{tr}\left( a\cdot \bar{b}\right )$ where $\bar{b}$ is the conjugate of $b$ and $a\cdot \bar{b}$ is the Hermitian inner product of $a$ and $b$. The trace operator is computed as $\text{tr}(\omega)= \text{tr}(\bar{\omega})=1$ and $\text{tr}(0)=\text{tr}(1)=0$.
	
	The matrix $\vec{S}$ is called the \emph{check matrix} and every row of $\vec{S}$ is called a \emph{check}. The checks correspond to the $m$ stabilizers which generate the stabilizer group $\mathcal{S}$. An $[[n,k,d]]$ stabilizer code is a $2^k$-dimensional subspace of the $n$-qubit Hilbert space $(\mathbb{C}^2)^{\otimes n}$ defined as the common +1 eigenspace of $\mathcal{S}$. Additionally, we define $\mathcal{N}(\mathcal{S})$ to be the normalizer of $\mathcal{S}$ and $\vec{S}^{\perp}$ to be the matrix containing the vectors corresponding to the $2n-m$ generators of $\mathcal{N}(\mathcal{S})$. The minimum distance $d$ of a \ac{QSC} is defined as the lowest error weight in $\mathcal{N}(\mathcal{S})\backslash \mathcal{S}$. A code is called \emph{degenerate} if $\mathcal{S}$ contains a stabilizer of weight less than $d$. Moreover, a code is highly degenerate if $d$ is much larger than the minimum weight of the stabilizers. 
	\vspace{-1em}
	
	\subsection{Codes Used in this Study}
	Two families of codes are considered in this work to evaluate the proposed decoder, namely \ac{GB} codes~\cite{panteleev2021degenerate} and toric codes~\cite{kitaev2006anyons}. 
	
	\Ac{GB} codes are constructed by choosing
	$\vec{H}_X' = \begin{pmatrix}
		\vec{A}&\vec{B}\\
	\end{pmatrix}$ and $\vec{H}_Z' = \begin{pmatrix}
		\trans{\vec{B}}&\trans{\vec{A}}\\
	\end{pmatrix}$
	where $\vec{A}$ and $\vec{B}$ are $\frac{n}{2}\times \frac{n}{2}$ square circulant matrices of rank $m/2$. For constructing the check matrix, $m/2$ linearly independent rows are chosen from both $\vec{A}$ or $\vec{B}$. Therefore, \ac{GB} codes naturally possess an overcomplete set of checks.

	An $[[n=2d^2,k=2,d]]$ toric code is a topological code constructed from a $d\times d$ lattice embedded on the surface of a torus. The qubits are placed on the edges of the lattice. The vertex and plaquette operators define the $\vec{X}$ and $\vec{Z}$ stabilizers, respectively. A toric code has $d^2$ $\vec{X}$ stabilizers of weight $4$ from the vertex operators and $d^2$ $\vec{Z}$ stabilizers of weight $4$ from the plaquette operators. Among them, one $\vec{X}$ stabilizer and one $\vec{Z}$ stabilizer are redundant.
	
	The two code families are connected by the lifted product construction~\cite{panteleev2021quantum} while having some contrasting properties. The \ac{GB} codes considered in this work usually have a larger minimal distance than the considered toric codes. However, the GB codes are not degenerate while the toric codes contain many low-weight stabilizers which lead to degenerate errors. 
	
	\begin{figure}
		\includegraphics{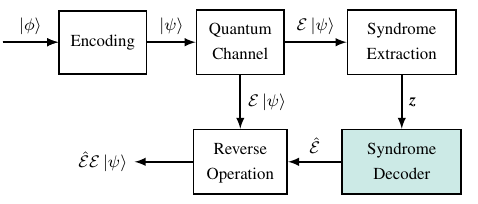}
		\caption{Block diagramm of \ac{QEC} using \acp{QSC}.}
		\label{fig:QECblock}
	\end{figure}

	\subsection{Error Correction with Stabilizer Codes}
	The workflow of error correction with stabilizer codes is depicted in Fig.~\ref{fig:QECblock}. To protect a logical quantum state $\ket{\phi}$ with $k$ qubits, it is encoded to a state $\ket{\psi}$ of $n$ qubits using a stabilizer code. The noise in the quantum memory can be modeled as a depolarizing channel with depolarizing probability $\epsilon$. In this channel, the errors $\vec{X},\vec{Z}$ and $\vec{Y}$ occur equally likely with probability $\frac{\epsilon}{3}$. For a potentially corrupted state, the syndrome vector $\vec{z}$ is extracted and sent to a decoder which aims to find the most probable error given the syndrome $\vec{z}$. The focus of this work is to improve the syndrome decoder, which is a \ac{BP4} decoder. The decoding result is used as a reverse operator and applied to the corrupted quantum state. This will hopefully recover the state $\ket{\psi}$. To be more precise: Let $\hat{\vec{e}}$ be the estimate of $\vec{e}$ by the decoder and $\hat{\vec{\mathcal{E}}}$ be its corresponding Pauli error. The decoder aims to find an $\hat{\vec{e}}$ yielding the same syndrome $\vec{z}$ and fulfilling $\hat{\vec{\mathcal{E}}}\vec{\mathcal{E}}\in \mathcal{S}$. The latter can be checked by
	\begin{equation}
		\label{eq:correction}
		\langle (\vec{e}+\hat{\vec{e}}), \vec{S}_i^{\perp}\rangle=0
	\end{equation}
	for every row $i\in \{1,2,\ldots, 2n-m\}$ of $\vec{S}^{\perp}$. 
	
	From \eqref{eq:correction}, we can see that when correcting errors using a \ac{QSC}, four outcomes may happen:
	\begin{itemize}
		\item Type I success: the estimated error is exactly the error that occurred, i.e., $\vec{e}+\hat{\vec{e}} = \vec{0}$.
		\item Type II success (with degeneracy): the estimated error is not exactly the same as $\vec{e}$ but \eqref{eq:correction} holds.
		\item Type I failure (flagged): the \ac{BP} decoder is not able to find an $\hat{\vec{e}}$  that matches the syndrome $\vec{z}$.
		\item Type II failure (unflagged): the syndrome matches but~\eqref{eq:correction} does not hold. This will lead to an undetectable erroneous quantum state.
	\end{itemize}

	One way to improve the decoding of \acp{QSC} is by designing a decoder that maximizes the probability of type I success, following classical decoding principles. However, this approach often falls short in achieving satisfactory decoding performance for \ac{QSC}s. To obtain a near-optimal decoder for quantum codes, it is crucial to exploit degeneracy and improve type I and II success rates jointly. This motivates the utilization of \ac{NN} decoders since there is no good decoding algorithm that systematically exploits degeneracy so far.

	\section{Belief Propagation Decoder}
	\label{sec:BP}
\begin{figure}
		\centering
		\includegraphics{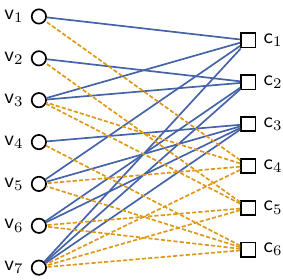}
		\caption{Tanner graph of the $[[7,1,3]]$ quantum BCH code, with VNs represented by circles and CNs by squares. Blue solid edges represent coefficients $\omega$ ($\bm{X}$ type), while yellow dashed edges represent coefficients~$\overline{\omega}$ ($\bm{Z}$ type).}
		\label{fig:tanner}
	\end{figure}

	BP decoding of QLDPC codes is performed on the \emph{Tanner graph} associated with the code. The Tanner graph is a bipartite graph with two classes of vertices. The first class of vertices are the \acp{VN}, each corresponding to a code bit (or qubit) and thus to a column of the check matrix $\vec{S}$. The second class of vertices are the \acp{CN}, each corresponding to a check and thus to a row of the check matrix. A VN $\mathsf{v}_i$ is connected to a CN $\mathsf{c}_j$ if the corresponding entry $S_{j,i}\neq 0$, where $S_{j,i} \in$ GF(4)$\backslash \{0\}$ is called the coefficient of the edge.
	For example, consider the $[[7,1,3]]$ quantum \ac{BCH} code with both $\vec{H}_X'$ and $\vec{H}_Z'$ being the \ac{PCM} of a classical $[7,4,3]$ \ac{BCH} code:
	\begin{equation*}
		\Vec{H}_{\text{BCH}} = \begin{pmatrix}
			1&0&1&0&1&0&1\\
			0&1&1&0&0&1&1\\
			0&0&0&1&1&1&1\\
		\end{pmatrix}.
	\end{equation*}
	The check matrix $\vec{S}\in $GF(4)$^{6\times 7}$ is obtained as
	\begin{equation}
		\vec{S} = \begin{pmatrix}
			\omega \vec{H}_{\text{BCH}}    \\
			\bar{\omega}\vec{H}_{\text{BCH}}
		\end{pmatrix}.
		\label{eq:BCH}
	\end{equation}
	The Tanner graph of this code is depicted in Fig.~\ref{fig:tanner}.

	In this work, the syndrome decoder in Fig.~\ref{fig:QECblock} is the quaternary \ac{BP4} decoder~\cite{DM98,DF07,lai2021log}, as binary decoding ignores the correlation between $\bm{X}$ and $\bm{Z}$ type of errors and is thus sub-optimal and has a higher error floor compared with a \ac{BP4} decoder, which takes the correlation between $\vec{X}$ and $\vec{Z}$ errors into account~\cite{panteleev2021degenerate,lai2021log}. The main disadvantage of the conventional BP4 decoder is the high complexity due to the passing of vector messages instead of scalar messages as in \ac{BP2}. The problem is solved by the recently proposed refined BP4 decoder with scalar messages~\cite{kuo2020refined,lai2021log}.
	
	\subsection{Refined BP4 Decoder}
	
	In this work, we use the log-domain refined BP4 decoder proposed in~\cite{lai2021log}. It exploits the fact that the syndrome of a stabilizer code is binary, indicating whether the error commutes with the stabilizer or not, enabling scalar message passing. Next, we briefly review the algorithm. 
	
	For every \ac{VN} $\mathsf{v}_i$, where $ i\in$ $\{1,2,\ldots, n\}$, we initialize the \ac{LLR} vector $\vec{\Gamma}_{i\to j}$ as $\vec{\Lambda}_i=\begin{pmatrix}
		\Lambda_i^{(1)}&\Lambda_i^{(\omega)}&\Lambda_i^{(\bar{\omega})}\\
	\end{pmatrix}\in \mathbb{R}^3$ with
	\[
	\Lambda_i ^{(\zeta)} = \ln \left(\frac{P(e_i=0)}{P(e_i=\zeta)}\right)=\ln \left(\frac{1-\epsilon_0}{\frac{\epsilon_0}{3}}\right),
	\]
	where $\zeta\in$ GF(4)$\backslash \{0\}$ and $\epsilon_0$ is the estimated physical error probability of the channel. To exchange scalar messages, a \textit{belief-quantization operator} $\lambda_{\eta}: \mathbb{R}^3 \to \mathbb{R}$ is defined as
	\begin{align*}
		\lambda_{\eta}(\vec{\Lambda}_i) &= \ln \left(\frac{P(\langle e_i, \eta\rangle=0)}{P(\langle e_i, \eta\rangle=1)}\right)\\
		&=\ln  \left(\frac{1+\me^{-\Lambda^{(\eta)}_i}}{\sum_{\zeta\neq 0, \zeta \neq \eta}\me^{-\Lambda^{(\zeta)}_i}}\right).
	\end{align*}
	
	The operator $\lambda_{\eta}$ maps the \ac{LLR} vector onto a scalar \ac{LLR} of the binary random variable $\langle e_i,\eta \rangle$ where $\eta$ runs over the nonzero entries of $\vec{S}$. The initial scalar \ac{VN} messages are calculated as 
	\begin{equation}
		\lambda_{i\to j} :=\lambda_{S_{j,i}}(\vec{\Gamma}_{i\to j})
		\label{eq:VNinitial}
	\end{equation}
	and are passed to the neighboring \acp{CN} where $i\to j$ denotes the message from VN $\mathsf{v}_i$ to CN~$\mathsf{c}_j$.
	\begin{figure*}
		\centering
		\includegraphics{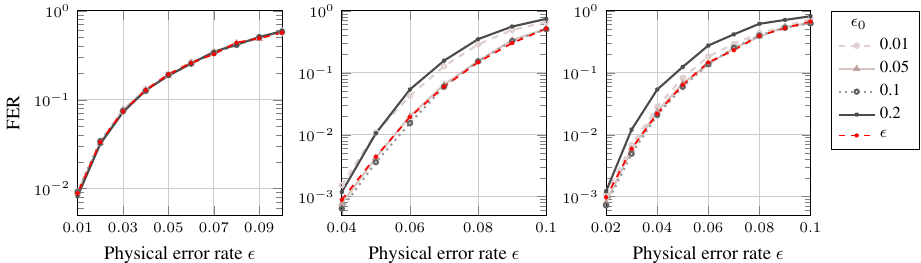}
		\caption{FER under different initial $\epsilon_0$ values: Left figure: toric code ($d=8$), middle figure: GB-A1 code, right figure: GB-A2 code.}
		\label{fig:ep0search_L8}
	\end{figure*}

	\begin{table*}[t]
		\centering
		\begin{tabular}{|clccccccc}
			\toprule
			\multicolumn{2}{c}{Code}                          & $n$   & $k$  & $d$  & $w_c$ &$m_{\text{oc}}$& $\epsilon_0$ (OBP4) & $(\epsilon_0,\,w_\mathrm{r})$ (OBP4-opt)\\
			\midrule
			\multicolumn{1}{r}{\multirow{4}{*}{GB}}    & A1   & 254 & 28 & 14-20 & 10  &254 &0.1&-\\
			\multicolumn{1}{c}{}                       & A2   & 126 & 28 & 8     & 10  &126 &0.1&-\\
			\multicolumn{1}{c}{}                       & A3   & 48  & 6  & 8     & 8   &2000 &0.3&-\\
			\multicolumn{1}{c}{}                       & A4   & 46  & 2  & 9     & 8   &800 &0.3&-\\
			
			\hline
			\multicolumn{1}{r}{\multirow{4}{*}{toric}} & $d=4$  & 32  & 2  & 4     & 4  &96 &0.45&-\\
			\multicolumn{1}{c}{}                       & $d=6$  & 72  & 2  & 6     & 4  &216 &0.48& (0.35, 0.1 ) \\
			\multicolumn{1}{c}{}                       & $d=8$  & 128 & 2  & 8     & 4  & 384& 0.49 & (0.37, 0.1)   \\
			\multicolumn{1}{c}{}                       & $d=10$ & 200 & 2  & 10    & 4  & 600& 0.48& (0.45, 0.15)  \\
			\bottomrule
		\end{tabular}
		\caption{Parameters of the codes used to evaluate the proposed decoding algorithms. Parameter $w_c$ is the CN degree of the original check matrix and $m_{\text{oc}}$ is the number of rows of the overcomplete check matrix. The optimal values of $\epsilon_0$ for OBP4 and $(\epsilon_0,\,w_\mathrm{r})$ for OBP4-opt used to obtain the results in Fig.~\ref{fig:FER_toric_OBP} are listed on the right side.}
		\vspace{-1em}
		\label{tab:ini_para}
	\end{table*}
	The outgoing messages of \ac{CN} $\mathsf{c}_j$, $j\in \{1,2,\ldots, m\}$,  are calculated using
	\begin{equation}
		\label{eq:CNupdate}
		\Delta_{i\lto j} = (-1)^{z_j}\cdot \underset{i'\in \mathcal{N}(j)\backslash\{i\}}{\boxplus} \lambda_{i'\to j},
	\end{equation} where $\mathcal{N}(j)$ denotes the indices of the neighboring VNs of $\mathsf{c}_j$ and the $\boxplus$ operation is defined as
	\[
	\overset{I}{\underset{i=1}{\boxplus}} x_i \coloneqq 2\tanh^{-1}\left( {\prod_{i=1}^{I}} \tanh\frac{x_i}{2} \right).
	\]
	
	At the \ac{VN} update, we first calculate the \ac{LLR} vector $\vec{\Gamma}_{i\to j}=\begin{pmatrix}
		{\Gamma}_{i\to j}^{(1)}&{\Gamma}_{i\to j}^{(\omega)}&{\Gamma}_{i\to j}^{(\bar{\omega})}\\
	\end{pmatrix}$ with
	\begin{equation}
		\label{eq:VNupdate}
		\Gamma^{(\zeta)}_{i\to j} = \Lambda_i^{(\zeta)} + \sum_{\substack{j'\in \mathcal{M}(i)\backslash\{j\}, \\ \langle \zeta, S_{j',i}\rangle=1}}\Delta_{i\lto j'},
	\end{equation}
	for all $\zeta\in$ GF(4)$\backslash \{0\}$ with $\mathcal{M}(i)$ denoting the indices of the neighboring CNs of VN $\mathsf{v}_i$. Then the outgoing messages $\lambda_{i\to j} = \lambda_{S_{j,i}}(\vec{\Gamma}_{i\to j})$ are calculated and passed to the neighboring CNs. 
	
	To estimate the error, a hard decision is performed at the \ac{VN}s by calculating $\vec{\Gamma}_i$ for $i\in \{1,2,\ldots, n\}$ with
	
	\begin{equation}
		\label{eq:hard-decision}
		\Gamma^{(\zeta)}_{i} = \Lambda_i^{(\zeta)} + \sum_{\substack{j\in \mathcal{M}(i)
				\\ \langle \zeta, S_{j,i}\rangle=1}} \Delta_{i\lto j},
	\end{equation}
	for all $\zeta\in$ GF(4)$\backslash \{0\}$. If all $\Gamma^{(\zeta)}_{i}>0$, then $\hat{e}_i=0$, otherwise $\hat{e}_i=\text{argmin}_{\zeta} \Gamma^{(\zeta)}_{i}$.
	
	The iterative process is performed until the maximum number of iterations $L$ is reached or the syndrome is matched.
	\vspace{-1em}

	\subsection{Initialization of the BP Decoder}
	
	During our experiments, we noticed that often a certain fixed initialization of $\epsilon_0$ tends to provide the best decoding performance for a wide range of physical error probabilities $\epsilon$. This is overlooked in many previous works, where $\epsilon_0$ is simply taken as the physical channel probability $\epsilon$.
	
	To explain this phenomenon, we look at the concept of a correctable error set proposed in~\cite{manabu2012} in the study of classical binary symmetric channels (BSCs). For a given \ac{QLDPC} code with a fixed Tanner graph, the set of correctable errors of a \ac{BP} decoder depends solely on the initialization value $\epsilon_0$. Depending on the weight distribution of the correctable error set, the optimal choice of $\epsilon_0$ is related to $\epsilon$ but it is often a distinct value~\cite{manabu2012}.
	
	To illustrate this point, we present in Fig.~\ref{fig:ep0search_L8} the BP4 decoding results for the toric code with $d=8$ and the GB codes A1 and A2 for different values of $\epsilon_0$. The code parameters are given in Tab.~\ref{tab:ini_para} and the construction matrices for the GB codes are given in~\cite[Appendix B]{panteleev2021degenerate}.  For the toric code (leftmost figure), no difference is seen for different $\epsilon_0$. This is due to the very low success rate of plain \ac{BP} decoding for toric codes, regardless of the chosen $\epsilon_0$.
	
	However, when we examine the decoding results for the \ac{GB} A1 code (middle figure) and A2 code (rightmost figure), it becomes evident that different $\epsilon_0$ values yield distinct decoding performances and choosing $\epsilon_0 = \epsilon$ leads to sub-optimal decoding performance. Similar results are observed for codes with other parameters as well. Therefore, to determine a good decoder configuration, a line search or another optimization method should be conducted to identify the best $\epsilon_0$ value.
	
	Furthermore, in what follows, the significance of using different initial $\epsilon_0$ values becomes more pronounced when redundant checks are introduced.

	\section{Overcomplete Check Matrices}
	\label{sec:overcomplete}
	Overcomplete check matrices refer to check matrices with redundant rows in addition to the original full-rank check matrix. \Ac{BP4} can be performed on the Tanner graph associated with the overcomplete check matrices. We call this decoder \emph{\ac{OBP4}}. In this section, we give two possible approaches to construct overcomplete check matrices and then investigate the reason for performance improvement from \ac{OBP4}.
	
	\subsection{Constructing Redundant Checks}
	\label{sec:overcomplete_construct}
	\begin{figure}
		\centering
		\includegraphics{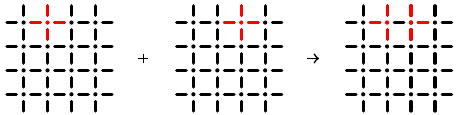}
		\caption{Example of the construction of a weight-6 stabilizer for the toric code with $d=4$. The red edges are the qubits which support an $X$ stabilizer.}
		\label{fig:toric_redundant_check}
	\end{figure}
	To construct redundant checks, we treat $\vec{H}_X$ and $\vec{H}_Z$ as two separate binary matrices. Then, constructing overcomplete check matrices is identical to finding a large number of low-weight stabilizers. Two approaches are used in this work. The first one is based on an exhaustive search, which is generally intractable. However, it is possible to use probabilistic approaches~\cite{Leon88} to increase the chance of finding low-weight stabilizers in the exhaustive search. This approach works well for the short-length codes considered in this work. 
	
	The second approach relies on the topological structure of topological codes. For example, as depicted in Fig.~\ref{fig:toric_redundant_check}, combining two neighboring stabilizers in a toric code yields a new weight-6 stabilizer. This results from each toric code stabilizer being supported by four qubits, with adjacent stabilizers overlapping by one qubit. Consequently, we can construct $2n$ redundant weight-6 $\vec{X}$ stabilizers and $2n$ redundant weight-6 $\vec{Z}$ stabilizers for every toric code. Combined with the $n$ weight-4 stabilizers, we obtain an overcomplete check matrix of size $3n\times n$ for toric codes, which is used in this work.
	
	The value of the redundant syndromes can be obtained without additional syndrome extraction operations. Let $\vec{H}$ be either $\vec{H}_X$ or $\vec{H}_Z$. The \ac{PCM} $\vec{H_{\text{oc}}}$ with redundant rows is obtained by $\vec{H_{\text{oc}}} = \vec{MH}$ with $\vec{M}$ being a binary matrix of size $m_{\text{oc}}\times m$. 
	The original syndrome is calculated as $\vec{H}\trans{\vec{e}} = \vec{z}$. Then, the new syndrome associated with $\vec{H}_{\text{oc}}$ is given by
	\[\vec{z}_{\text{oc}} = \vec{H}_{\text{oc}}\trans{\vec{e}} =\vec{MH}\trans{\vec{e}} = \vec{M}\vec{z}, \]
	being a linear mapping of $\vec{z}$ represented by $\vec{M}$.

	\subsection{Heuristic Analysis}
	In this section, we present the heuristics which inspired our use of overcomplete check matrices.
	
	The idea of performing \ac{BP} decoding over an overcomplete parity-check matrix has been investigated in~\cite{lian2019learned,halford2006random,bossert1986hard,kothiyal2005iterative,Jiang2006iterative,buchberger2020pruning} for classical linear codes. One of the initial motivations was to perform more node updates in parallel to reduce the effect of short cycles. It is interesting that this method is able to improve the decoding performance of \ac{QLDPC} codes significantly, both in the case of a limited as well as an unlimited number of iterations. This differs from decoding classical linear codes where using overcomplete check matrices only improves convergence speed and does not show obvious gains when a large number of iterations is used. 
	
	For \ac{QLDPC} codes, this approach resembles matrix augmentation~\cite{rigby2019modified}, where a fraction of the rows of the check matrix are duplicated. The advantage is that the messages associated with the duplicated check node are magnified which helps in breaking the symmetry during decoding and leading to a (hopefully) correct error estimate~\cite{rigby2019modified}. 
	
	We demonstrate an extra benefit of the proposed method with a toy example of the $[[7,1,3]]$ quantum \ac{BCH} code described in Sec.~\ref{sec:BP}, whose check matrix is given in \eqref{eq:BCH}.
	
	Consider the error $\vec{\mathcal{E}} = \vec{Y}_7$. The syndrome $\vec{z}$ of $\vec{e}$ is
	\[
	\vec{z} = \begin{pmatrix}
		1\;1\;1\;1\;1\;1\\
	\end{pmatrix}.
	\]
	Assume that we initialize the decoder with $\epsilon_0=0.1$. According to \eqref{eq:VNinitial}, all the initial \ac{VN} messages are $2.64$. Then, in the first \ac{CN} update, all \ac{CN} messages are $-1.55$ according to \eqref{eq:CNupdate}. Based on these messages, in the next hard decision step, the decoder estimates the error as $\hat{\vec{\mathcal{E}}}=\vec{Y}_3\vec{Y}_5\vec{Y}_6\vec{Y}_7$ which produces the same syndrome as $\vec{z}$. However, \eqref{eq:correction} does not hold and  we end up with an unflagged error.

\begin{figure}[tb]
		\centering
		\includegraphics{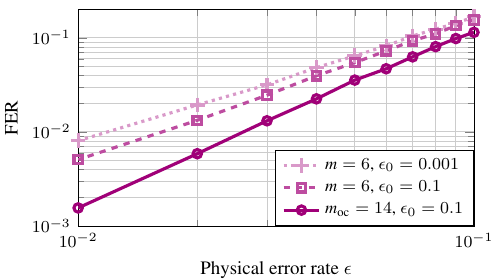}
		
		\caption{BP decoding results for the [[7,1,3]] CSS code with original and overcomplete check matrix ($L=32$).}
		\label{fig:FER71}
	\end{figure}

	In this example, the decoder wrongly estimates the error because all CNs have a syndrome value of $1$. This is a strong indication that the error has a high weight. Except for \acp{VN} $\mathsf{v}_1$, $\mathsf{v}_2$, and $\mathsf{v}_4$ with degree 2, all VNs that are connected to more than 2 \acp{CN} are erroneously estimated in the first hard-decision step. Duplicating some rows of the check matrix does not help in this case.\footnote{However, note that this problem could be solved by initializing the decoder with a very small $\epsilon_0$ such as $0.001$ at the cost of degrading the overall decoding performance (depicted in Fig.~\ref{fig:FER71}), as in most cases, the decoder has a tendency towards trivial errors.}
	
	Now we use the overcomplete $\vec{H}_X$ and $\vec{H}_Z$ with 7 rows, i.e., we take all the linear combinations of the $3$ rows of the original $\vec{H}_{\text{BCH}}$ except the trivial case (all-zero vector). The new syndrome is
	\[
	\vec{z}_{\text{oc}} = \begin{pmatrix}
		1\;   1 \;  0 \;  1  \; 0\;  0 \;  1 \;  1 \;  1 \;  0 \;  1  \; 0 \;  0 \;  1\\
	\end{pmatrix}.
	\]
	Although $\vec{z}_{\text{oc}}$ contains the same amount of information as $\vec{z}$ with respect to the coset where the error $\vec{\mathcal{E}}$ is located, the former is easier for the decoder to interpret. In the first hard-decision step, the error is correctly estimated. Figure~\ref{fig:FER71} depicts the \ac{FER} curves corresponding to decoding using the overcomplete check matrix and the original check matrix for different $\epsilon_0$.
	\section{Neural Belief Propagation}
	\label{sec:NBP}
	\Ac{NBP} has been shown to be effective in improving the decoding performance of classical linear codes without increasing the decoding latency, see, e.g.,~\cite{nachmani2016learning,nachmani2018deep,lian2019learned}. Using \ac{NBP} to enhance the decoding of quantum stabilizer codes has been investigated in~\cite{liu2019neural,xiao2019neural} for the \ac{BP2} decoder, which decodes $\vec{X}$ and $\vec{Z}$ errors separately. In this section, we propose to extend the \ac{NBP} concept to the \ac{BP4} decoder with an adapted loss function tailored to exploit degeneracy. The proposed \ac{NBP} model can be applied to both the original check matrix as well as the overcomplete check matrix, resulting in an \emph{\ac{NBP4}} decoder and an \emph{\ac{NOBP4}} decoder, respectively.
	
	\subsection{Framework}
	An \ac{NBP} decoder is an \ac{NN} decoder obtained by unfolding a \ac{BP} decoder over the decoding iterations~\cite{nachmani2016learning}. The \ac{BP} decoder can be based on the original Tanner graph of the \ac{QLDPC} code or the Tanner graph associated with an overcomplete check matrix. The \acp{VN} and \acp{CN} are interpreted as neurons of an \ac{NN} and the activation functions of the \ac{NN} describe the node update equations. Trainable weights are added to the messages passed along the edges of the graph, i.e., the update rules \eqref{eq:CNupdate} and \eqref{eq:VNupdate} become
	
	\begin{equation}
		\Delta_{i\lto j} = (-1)^{z_j}\cdot \underset{i'\in \mathcal{N}(j)\backslash\{i\}}{\boxplus}  w_{\mathsf{v},i',j}^{(\ell)}\cdot \lambda_{i'\to j}
		\label{eq:NBP_CN}
	\end{equation}
	
	and
	\begin{equation}
		\Gamma^{(\zeta)}_{i\to j} = w_{\mathsf{v},i}^{(\ell)}\Lambda_i^{(\zeta)} + \sum_{\substack{j'\in \mathcal{M}(i)\backslash\{j\}, \\ \langle \zeta, S_{j'i}\rangle=1}}w_{\mathsf{c},i,j'}^{(\ell)}\cdot \Delta_{i\lto j'},
		\label{eq:NBP_VN}
	\end{equation}
	where $\ell$ is the decoding iteration index and $w_{\mathsf{v},i',j}^{(\ell)}$, $w_{\mathsf{c},i,j'}^{(\ell)}$ and $w_{\mathsf{v},i}^{(\ell)}$ denote the real-valued trainable weights applied to the VN messages, CN messages, and channel \ac{LLR} values, respectively.
	
	\subsection{Loss Function}
	
	In \cite{liu2019neural}, a loss function for BP2 decoding has been proposed which takes degeneracy into account.
	We extend this loss function to the BP4 case. Using $\vec{\Gamma}_{i}$ calculated by \eqref{eq:hard-decision} in the hard-decision step, we compute
	\begin{equation}
		\label{eq:loss_aid}
		P(\langle \hat{e}_i, \eta \rangle=1|\vec{z}) = \left(1+\me^{-\lambda_{\eta}(\vec{\Gamma}_i)}\right)^{-1}
	\end{equation}
	for $\eta \in$ GF(4)$\backslash \{0\}$, denoting the estimated probability of the $i$-th error commuting with $\vec{X}$,  $\vec{Z}$, and $\vec{Y}$, respectively. Then, the proposed per-error pattern loss function can be written as
	\begin{equation}
		\label{eq:loss}
		\mathcal{L}(\vec{\Gamma};\vec{e}) = \sum_{j=1}^{2n-m} f\left( \sum_{i=1}^{n} P\left(\langle e_i+\hat{e}_i, S_{j,i}^{\perp} \rangle =1|\vec{z}\right)\right)
	\end{equation}
	where $f(x)=|\sin(\pi x/2)|$, which approaches $0$ with $x$ approaching any even number (i.e, it realizes a ``soft modulo 2'' function)~\cite{liu2019neural}. The loss value is summed up over all rows $\bm{S}^{\perp}_j$ of $\bm{S}^{\perp}$. For each row $j$, we sum up the values $P\left(\langle e_i+\hat{e}_i, S_{j,i}^{\perp} \rangle =1|\vec{z}\right)$ for all the elements $S_{j,i}^{\perp}$ in $\bm{S}^{\perp}_j$, representing the probability of $S_{j,i}^{\perp}$ being unsatisfied after estimating $e_i$ as $\hat{e}_i$. It can be calculated as $P\left(\langle \hat{e}_i, S_{j,i}^{\perp} \rangle =1 +\langle e_i,  S_{j,i}^{\perp} \rangle|\vec{z}\right)$. If $\langle e_i,  S_{j,i}^{\perp} \rangle =0$, we directly calculate $P\left(\langle \hat{e}_i, S_{j,i}^{\perp} \rangle =1|\vec{z}\right)$ using \eqref{eq:loss_aid}, with $\eta$ being $S_{j,i}^{\perp}$. In contrast, when $\langle e_i,  S_{j,i}^{\perp} \rangle =1$, we calculate $P\left(\langle \hat{e}_i, S_{j,i}^{\perp} \rangle =0|\vec{z}\right)=1-P\left(\langle \hat{e}_i, S_{j,i}^{\perp} \rangle =1|\vec{z}\right)$. The loss function \eqref{eq:loss} is minimized if \eqref{eq:correction} holds.

	\subsection{Initial Weights of the Redundant Checks}
	When a large number of redundant CNs is used, it becomes crucial to properly normalize the messages as the dependency between the messages becomes severe during decoding. Therefore, sometimes using a relatively large $\epsilon_0$ does not provide a sufficient degree of freedom for the normalization of the messages. Hence, we introduce another parameter $w_{\mathrm{r}}$, which is used as a normalization factor for the check node messages, similarly to the $w_{\mathsf{c},i,j}^{(\ell)}$ in \eqref{eq:NBP_VN}, but independent of the number of iterations $\ell$ and of the index of the connected CN $j$ and VN $i$, i.e., 
	\[
	\Gamma^{(\zeta)}_{i\to j} = \Lambda_i^{(\zeta)} + \sum_{\substack{j'\in \mathcal{M}(i)\backslash\{j\}, \\ \langle \zeta, S_{j'i}\rangle=1}}w_{\mathrm{r}}\cdot \Delta_{i\lto j'}.
	\]
	It can be used in the plain \ac{OBP4} decoder or as the initial weight of $w_{\mathsf{c},i,j'}^{(\ell)}$ for training an \ac{NOBP4} decoder. 
	
	When optimizing the decoder configuration, we simply set $w_{\mathrm{r}}=1$ if optimizing $\epsilon_0$ alone yields satisfactory decoding results, as it results in the best convergence speed. If this is not the case, a grid search is conducted to find the optimal pair $(\epsilon_0,\,w_\mathrm{r})$. This decoder is referred to as OBP4-opt decoder. In Appendix \ref{app:search}, details about the grid search for the optimal initial value pair $(\epsilon_0, w_{\mathrm{r}})$ are given. Table~\ref{tab:ini_para} summarizes the parameters of the exemplary codes and the corresponding decoder configurations used in this work.

	\subsection{Training}
	
	To train an NBP decoder, we employ the following procedure. First, we perform a line search to determine the optimal initial value of $\epsilon_0$ for the initial non-trained decoder. The initial values for $w_{\mathsf{v},i,j}^{(\ell)}$ and $w_{\mathsf{v},i}^{(\ell)}$ at the beginning of the training are set to 1. If an overcomplete check matrix is used, we also search for the optimal initial weight $w_{\mathrm{r}}$ for the redundant checks jointly with $\epsilon_0$. In this case, the initial value of $w_{\mathsf{c},i,j}^{(\ell)}$ will be $w_{\mathrm{r}}$. Otherwise, when no redundant checks are used, the initial value of $w_{\mathsf{c},i,j}^{(\ell)}$ is 1. Subsequently, the training of the \ac{NN} decoder is carried out using the proposed loss function \eqref{eq:loss} through plain stochastic gradient descent~\cite{goodfellow2016deep}.
	
	During training, mini-batches of 120 samples are used. Each mini-batch comprises six sets of 20 randomly generated error patterns from depolarizing channels with six distinct physical error probabilities $\epsilon$. For training the \ac{NBP4} decoder, the set of $\epsilon$ is $\{0.02, 0.03, \ldots, 0.07\}$, while for training the \ac{NOBP4} decoder, we use $\{0.06, 0.07, \ldots, 0.11\}$. The loss function \eqref{eq:loss} is evaluated after each decoding iteration $\ell$, resulting in a loss $\mathcal{L}^{(\ell)}$. The per-error pattern loss is determined as the minimum value across all iterations, i.e., as $\min \{\mathcal{L}^{(\ell)}: \ell \in \{1,2,\ldots, L\}\}$. In the case where multiple iterations yield the same value, we select the earliest iteration. The gradients are clipped to $10^{-3}$ to avoid exploding gradients in the first decoding iterations where the message magnitude is small.  Additionally, we gradually decrease the learning rate from 1 to 0.1 using a linear scheduler. For the \ac{NBP4} decoder, we train for $2000$ batches,  whereas for the \ac{NOBP4} decoder, we only train for $200$ batches.
	
	\REV{The trainable model parameters are summarized in Tab.~\ref{tab:trainable_parameter} and the hyper-parameters for training are summarized in Tab.~\ref{tab:hyper_parameter}.}

	\begin{table}[]
		\setlength{\tabcolsep}{3pt}
		\centering
		\caption{\REV{A summary of the trainable parameters.}}
		\REV{
			\begin{tabular}{c c}
				\toprule
				parameter& meaning \\
				\hline
				$w_{\mathsf{v},i,j}^{(\ell)}$ &  weight applied on message from $\mathsf{v}_i$ to $\mathsf{c}_j$ in the $\ell$-th iteration\\
				$w_{\mathsf{c},j,i}^{(\ell)}$ &  weight applied on message from $\mathsf{c}_j$ to $\mathsf{v}_i$ in the $\ell$-th iteration\\
				$w_{\mathsf{v},i}^{(\ell)}$ &  weight applied on the channel LLR in the $\ell$-th iteration\\
				\bottomrule
		\end{tabular}}
		
		\label{tab:trainable_parameter}
	\end{table}
	
	\begin{table}[]
		\setlength{\tabcolsep}{3pt}
		\centering
		\caption{\REV{A summary of the hyper-parameters.}}
		\REV{
			\begin{tabular}{c c c c c}
				\toprule
				model& \makecell{batch \\ size} & \makecell{batch \\ number} & \makecell{sampling \\ error rate $\epsilon$}&\makecell{learning\\rate}\\
				\hline
				NBP4&120&2000&$\{0.02, 0.03, \ldots, 0.07\}$&\makecell{$1$ to $0.1$ with \\linear scheduler}\\
				NOBP4&120&200&$\{0.06, 0.07, \ldots, 0.11\}$&\makecell{$1$ to $0.1$ with \\linear scheduler}\\
				\bottomrule
		\end{tabular}}
		\label{tab:hyper_parameter}
	\end{table}

	\begin{figure}[t]
		\centering
		\includegraphics{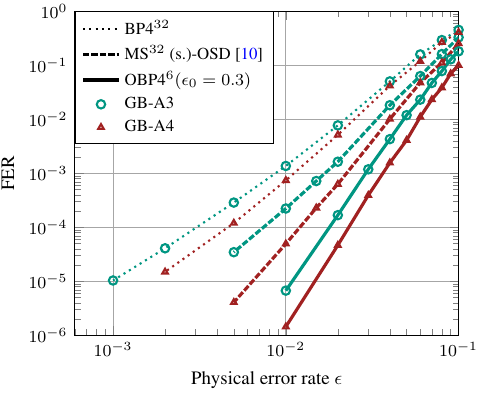}
		\caption{FER vs. depolarizing probability $\epsilon$ curves for the $[[48,6,8]]$ GB-A3 code and $[[46,2,9]]$ GB-A4 code. The reference curves are taken from~\cite{panteleev2021degenerate}, which use $32$ iterations of serial (s.) normalized \ac{MS} decoding concatenated with \ac{OSD}-10 post-processing.}
		\label{fig:FER_A3_A4}
		
	\end{figure}
	
	\begin{figure}[tb]  
		\centering
		\includegraphics{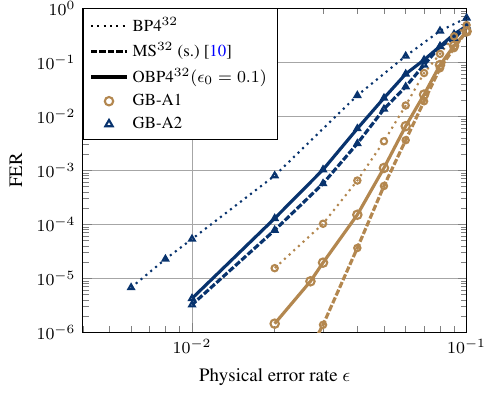}
		\caption{FER vs. depolarizing probability $\epsilon$ curves for the $[[254,28,d]]$ GB-A1 code and $[[126,28,8]]$ GB-A2 code. The reference curves are taken from~\cite{panteleev2021degenerate} where $32$ iterations of serial normalized \ac{MS} decoding is used.}
		\label{fig:FER_A1_A2}
	\end{figure}

	\begin{figure*}[t!]
		\centering
		\includegraphics{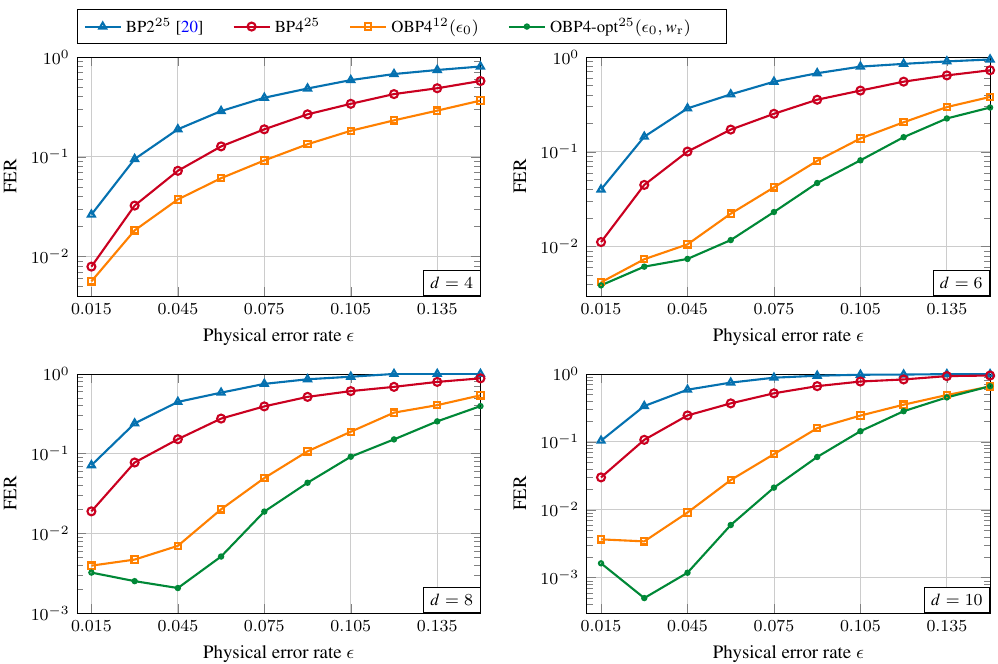}
		\caption{FER vs. depolarizing probability $\epsilon$ curves for the toric codes with $d\in\{4,6,8,10\}$ using plain \ac{BP}2 and \ac{BP}4 decoders. The BP2 results are taken from~\cite{liu2019neural}. The parameters used to produce the results for \ac{OBP4} and \ac{OBP4}-opt are listed in Tab.~\ref{tab:ini_para}.}
		\label{fig:FER_toric_OBP}
		\vspace{-1ex}
	\end{figure*}

	\subsection{Complexity}
	\REV{Firstly, the refined BP4 decoder has a computational complexity scaling linearly with the number of iterations and the number of edges in the Tanner graph, as one can see from \eqref{eq:VNupdate} and \eqref{eq:CNupdate}. Compared to BP2 decoders, the source of complexity increase is the belief-quantization step which can be implemented with complexity $\mathcal{O}(1)$ per edge~\cite{lai2021log}.
		Therefore, the complexity of the BP4 decoders is only slightly higher than the complexity of the BP2 decoders.}
	
	\REV{The proposed NBP4 decoders are based on the refined BP4 decoder. As shown in \eqref{eq:NBP_CN} and \eqref{eq:NBP_VN}, the proposed neural decoders require one additional multiplication with weights per edge compared to the refined BP4 decoder. Therefore, compared to conventional binary decoders, the complexity increase of the proposed NBP4 decoders is moderate.}
	
	\REV{Using an overcomplete check matrix increases the number of CNs by a factor of 1.5 to 10 compared to a full-rank matrix, based on the example codes presented. Additionally, while the computational complexity per CN remains unchanged, VNs require more additions. However, the number of required iterations is significantly reduced, often to half or even one-tenth, when using overcomplete matrices. Generally, we can decrease the maximum number of iterations as the number of redundant CNs increases, preserving similar overall complexity in both cases. Note that for QEC, the decrease in the number of iterations is advantageous, as low-latency decoding is crucial due to the limited coherent time of the qubits.}
	
	\section{Numerical Results}
	\label{sec:results}
	We assess the performance of our proposed novel decoder on the codes described in Sec.~\ref{sec:preliminaries} using Monte Carlo simulations. Specifically, we use the GB A1-A4 codes and toric codes with $d\in\{4,6,8,10\}$. The code parameters used are provided in Tab.~\ref{tab:ini_para}. To ensure sufficiently accurate results, 300 frame errors are collected for each data point. Throughout this work, we use a flooding schedule where all VNs/CNs are updated in parallel in each decoding iteration. The initial $\epsilon_0$ is set to $0.1$ unless explicitly stated.
	
	First, we investigate the effects of overcomplete check matrices in Sec.~\ref{sec:res_overcomplete}. Later, in Sec.~\ref{sec:res_NBP}, we show performance improvements due to NBP.

	\begin{figure*}[t!]
		\centering
		\includegraphics{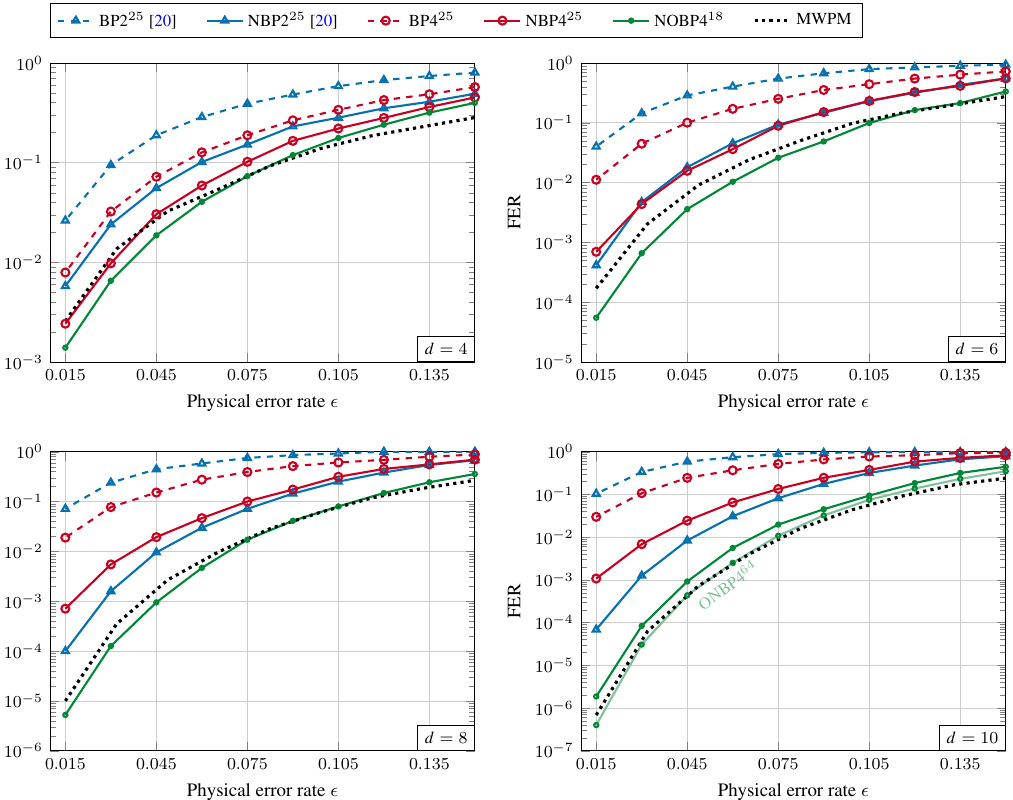}
		\caption{FER vs. depolarizing probability $\epsilon$ curves for the toric codes with $d\in\{4,6,8,10\}$ using the trained \ac{NBP} decoders.}
		\label{fig:FER_toric_NBP}
	\end{figure*}
	
	\subsection{Results with Overcomplete Check Matrices}
	\label{sec:res_overcomplete}
	\subsubsection{GB Codes}
	
	Figure~\ref{fig:FER_A3_A4} shows the decoding results using the original \ac{BP4} and the proposed \ac{OBP4} for the GB-A3 and GB-A4 codes. As described in Sec.~\ref{sec:overcomplete_construct}, the overcomplete check matrices are constructed using the algorithm from \cite{Leon88}. For the GB-A3 code, we construct the overcomplete check matrix using $48$ rows of weight $8$ and $1952$ rows of weight $12$. No checks of weight $10$ were found. For the GB-A4 code, we use $46$ rows of weight $8$ and $754$ rows of weight $10$. An improvement in post-decoding \ac{FER} by more than an order of magnitude is observed using the \ac{OBP4} decoder with only 6 decoding iterations, especially when the physical error rate is small. The proposed \ac{OBP4} decoder also outperforms the reference decoder from~\cite{panteleev2021degenerate}, which uses $32$ iterations of serial normalized \ac{MS} decoding concatenated with \ac{OSD}-10 post-processing.
	
	During our experiments, we observe that the \ac{OBP4} decoder performs especially well if the code possesses a large number of low-weight checks which can be used to construct overcomplete check matrices. For codes with only a small number of redundant low-weight stabilizers, a decoding performance gain is still observed but the gain is smaller when the ratio between the number of redundant rows and the block length is smaller. For example, in Fig.~\ref{fig:FER_A1_A2}, we show the decoding results for the GB-A1 and GB-A2 codes.  For both codes, the overcomplete check matrices consist of $n$ rows of weight 10. No further redundant rows are added, as both codes do not possess any other low-weight check with a weight less than 16. For both codes, the \ac{OBP4} decoder outperforms the original \ac{BP4} decoding and has only a small gap to the reference results which uses normalized \ac{MS} decoding with serial scheduling\cite{panteleev2021degenerate}.
	
	\subsubsection{Toric Codes}
	As depicted in Fig.~\ref{fig:FER_toric_OBP}, we also evaluate the decoding performance of BP4 and \ac{OBP4} decoding on toric codes with $d\in \{4,6,8,10\}$.  For comparison, we plot the \ac{BP2} decoding results taken from~\cite{liu2019neural} which were evaluated over the XZ channel where the $\vec{X}$ and $\vec{Z}$ errors are assumed to happen independently with a probability $\epsilon_{\mathrm{b}}$. For fairness, we scale the results from~\cite{liu2019neural} using $\epsilon_{\mathrm{b}}=\frac{2}{3}\epsilon$ as explained in~\cite{mackay2004sparse}.
	
	We see that for plain BP decoding, BP4 performs consistently better than BP2 decoding. This highlights the advantage of using a quaternary decoder. The \ac{BP4} decoding result can be further improved by constructing an overcomplete check matrix of size $3n\times n$ using the topological structure of a toric code as described in Sec.~\ref{sec:overcomplete_construct}. The performance improvement due to OBP4 is shown in Fig.~\ref{fig:FER_toric_OBP} (orange curves). For toric codes with $d\geq 6$, we also use the OBP-opt decoder where a second initial parameter $w_{\mathrm{r}}$ is used (green curves). For $d=4$, OBP-opt is not used as it does not show improvements compared to OBP.
	\REV{One can observe that the \ac{OBP4}-opt decoder exhibits an error floor, especially for the codes with relatively large block lengths where the post-decoding error rate does not decrease with the decreasing channel error rate. This is caused by our choice of initial parameters to ensure a good performance after training. We observed during optimization that we need to focus on parameters that yield good performance for medium to high physical error rates. The decoder tends to converge to high-weight errors and thus performs worse for low-weight errors, which are dominant when the physical error rate is low. We optimize at this working point as the error floor can later be mitigated by NBP and we observe that this configuration yields the best overall decoding performance across different physical error rates.}

	\subsection{Results for NBP}
	\label{sec:res_NBP}
	We apply both \ac{NBP4} and \ac{NOBP4} decoders to toric codes and compare the decoding results with the neural BP2 decoder from~\cite{liu2019neural} and the \ac{MWPM} decoder. The numerical results of the MWPM decoder are obtained via the Pymatching library~\cite{higgott2023sparse}. All results are plotted in Fig.~\ref{fig:FER_toric_NBP}.
	
	\subsubsection{FER results}
	We first compare the effect of NBP for both BP2 (red curves) and BP4 (blue curves) decoding without using any overcomplete check matrices. After training, both decoders improve the FER by orders of magnitude. This demonstrates the effectiveness of training. For toric codes with large $d$, the gain of NBP2 compared to BP2 is bigger than the gain of NBP4 compared to BP4, as NBP2 from~\cite{liu2019neural} also implements other enhancements such as residual connections and soft weights. However, the performance of the NBP2 decoder is lower bounded by the MWPM decoder, which is considered to be near optimum for the XZ channel. Figure~\ref{fig:FER_toric_NBP} shows that the decoding performance of the NOBP4$^{18}$ decoder is better than the \ac{MWPM} decoder except for $d=10$. We observe that for codes with large block lengths, more BP iterations (e.g., 64 iterations) are needed to achieve a good decoding performance. This is shown with the light green curve.
	
	Simulation results show that the \ac{NOBP4} decoder resolves both the error floor and convergence speed issues of the OBP4-opt decoder. Consequently, the NOBP4 decoder achieves better decoding performance and requires only 18 iterations, whereas the other decoders use 25 iterations.

	\subsubsection{Exploiting Degeneracy with \ac{NBP}}
	\begin{figure}[tb]
		\centering
		\includegraphics{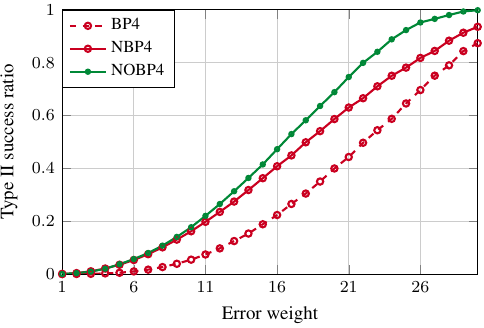}
		\caption{Fraction of type II success (success with degenerated errors) in relation to overall success plotted over the error weight for decoding the toric codes with $d=10$.}
		\label{fig:degeneracyToric10}
	\end{figure}
	In Fig.~\ref{fig:degeneracyToric10}, we depict the fraction of type II decoding successes in relation to the total number of decoding successes when decoding error patterns of weights ranging from 1 to 30 occur. By employing both \ac{NBP4} and \ac{NOBP4} decoders, the fraction of successful decodings with degenerate errors (type II success) exhibits a significant increase compared to the original \ac{BP4} decoder. This highlights the effectiveness of training in exploiting degeneracy, which explains the substantial improvement when decoding toric codes with neural decoders. Toric codes, particularly those with large minimum distance~$d$, possess many degenerate errors from their topological structure.
	
	Furthermore, We observe that the NN is trained to exploit degeneracy from the weights $w_{\mathsf{c},i,j'}^{(\ell)}$. When trained solely with low-weight errors, we observe a clear trend where low-weight checks tend to have high message weights after training. This trend maximizes the type I success rate by increasing the weights on low-weight checks, which aligns with observations in decoding classical codes where low-degree check nodes are advantageous. However, when trained with mixed errors of both low and high weights, this tendency diminishes. After training, it becomes sometimes more beneficial to converge to an error that differs from the actual error that occurred (type II success). This indicates that the \ac{NN} learns the degeneracy structure of the code.
	
	For \ac{GB} codes, the large minimum distance $d$ ensures that they almost always achieve a valid error estimate as long as the syndrome is matched. The fraction of type II errors is extremely small (typically below $10^{-3}$). Decoding failures mainly arise from the inability of the BP decoder to find an error estimate that matches the syndrome. Hence, enhancing the convergence of BP with overcomplete check matrices leads to improved performance.  However, since \ac{GB} codes are not highly degenerate, the type II success rate for \ac{GB} codes is also quite low (typically below $10^{-3}$). Training the \ac{NN} decoder slightly improves the type II success rate but also significantly increases the type II failure rate. Consequently, training the neural decoder does not yield further performance improvement for GB codes.

	\subsection{Summary of the Numerical Results}
	The proposed enhancements of the \ac{BP} decoder significantly improve the decoding performance for both \ac{GB} codes and toric codes. 
	However, the reasons for improvement are different. We observe that plain \ac{BP4} decoding performs significantly better on \ac{GB} codes than on toric codes. Furthermore, the decoding performance for both \ac{GB} and toric codes is further enhanced by constructing a suitable overcomplete check matrix. However, the training of \ac{NBP} decoding for GB code is not as successful as for toric codes, and thus, we omit results.

	\section{Conclusions and Outlook}
	\label{sec:conclusion}
	In this paper, we investigated two enhancements for the BP decoder of QLDPC codes, as well as their combination. The OBP4 decoder offers significant improvements in decoding performance and convergence speed. In contrast, the proposed NBP4 decoder effectively exploits degeneracy, but its improvement through training alone is limited without good initial parameters. By applying NBP to OBP4 decoding, we are able to address both of these disadvantages, resulting in superior decoding performance.
	
	The proposed decoder enhancements have several natural extensions that are not discussed in the scope of this paper. Decoding based on overcomplete check matrices can also be applied to other message-passing decoders such as the \ac{MS} decoder, which will yield similar decoding gains as for the BP decoder. The proposed loss function can also be used to train other model-based neural decoders such as a neural offset min-sum (NOMS) decoder~\cite{8006751} \REV{and other optimized min-sum decoders with efficient hardware implementation~\cite{ref1}\cite{ref2}}. As the proposed decoder reduces decoding latency, post-processing techniques could also be applied after \ac{BP} decoding. Additionally, considering circuit-level noise is also important for future work.
	
	\REV{The source code for this work is available at \url{https://github.com/kit-cel/Quantum-Neural-BP4-demo}. The trained model parameters are also included in the repository.} The provided Jupyter notebook can perform training and evaluation of the NBP model for arbitrary CSS with a given PCM. 
	
	\appendices    
	\begin{figure*}[t]
		\centering
		\includegraphics{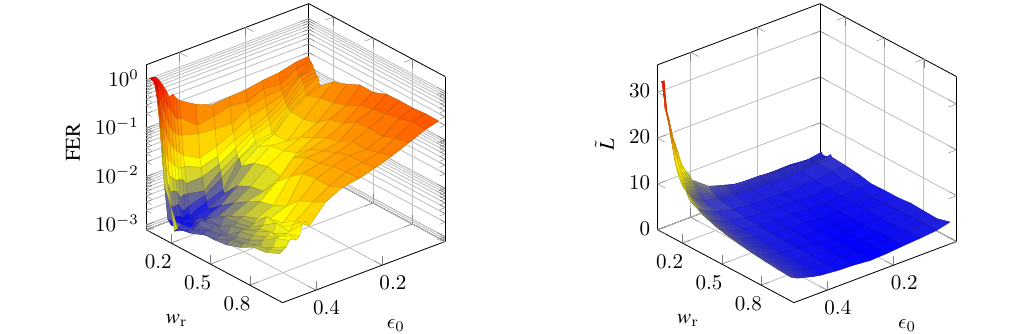}
		\caption{OBP4 decoding results for toric codes with $d=10$ under different configurations of ($\epsilon_0$, $w_\mathrm{r}$) over a depolarizing channel with $\epsilon=0.045$: Left figure shows the post decoding FER and the right figure shows the average number of BP decoding iterations until decoding success ($\Tilde{L}$).}
		\label{fig:ep0search_L10}
	\end{figure*}
	
	\section{Grid Search for the Initial Parameters}
	\label{app:search}
	Section~\ref{sec:BP} emphasizes the significance of the initial value $\epsilon_0$ in BP decoding, particularly when overcomplete check matrices are employed. To achieve the best decoding performance with a specific decoder, we conduct a line search to find the $\epsilon_0$ that minimizes the post-decoding FER. In some cases, optimizing over $\epsilon_0$ alone does not provide enough flexibility for optimization. Then we search for the optimal CN weight $w_{\mathrm{r}}$ for redundant checks jointly with $\epsilon_0$ via a grid search.
	
	Figure~\ref{fig:ep0search_L10} illustrates the post-decoding FER (left figure) and the average number of decoding iterations $\Tilde{L}$ until success (right figure) for the toric code with $d=10$. The left figure reveals a specific region where the decoder exhibits the best performance, characterized by a relatively large $\epsilon_0$ value and a small $w_{\mathrm{r}}$ value. However, the right figure shows that the number of iterations required for the BP decoder to converge increases with larger $\epsilon_0$ and smaller $w_\mathrm{r}$. Consequently, when the decoding performance is similar, we choose values that enable faster convergence. For other toric codes, the search results are similar to the ones shown in Fig.~\ref{fig:ep0search_L10} and are summarized in Tab.~\ref{tab:ini_para}.

	\begin{IEEEbiography}[{\includegraphics[width=1in,height=1.25in,clip,keepaspectratio]{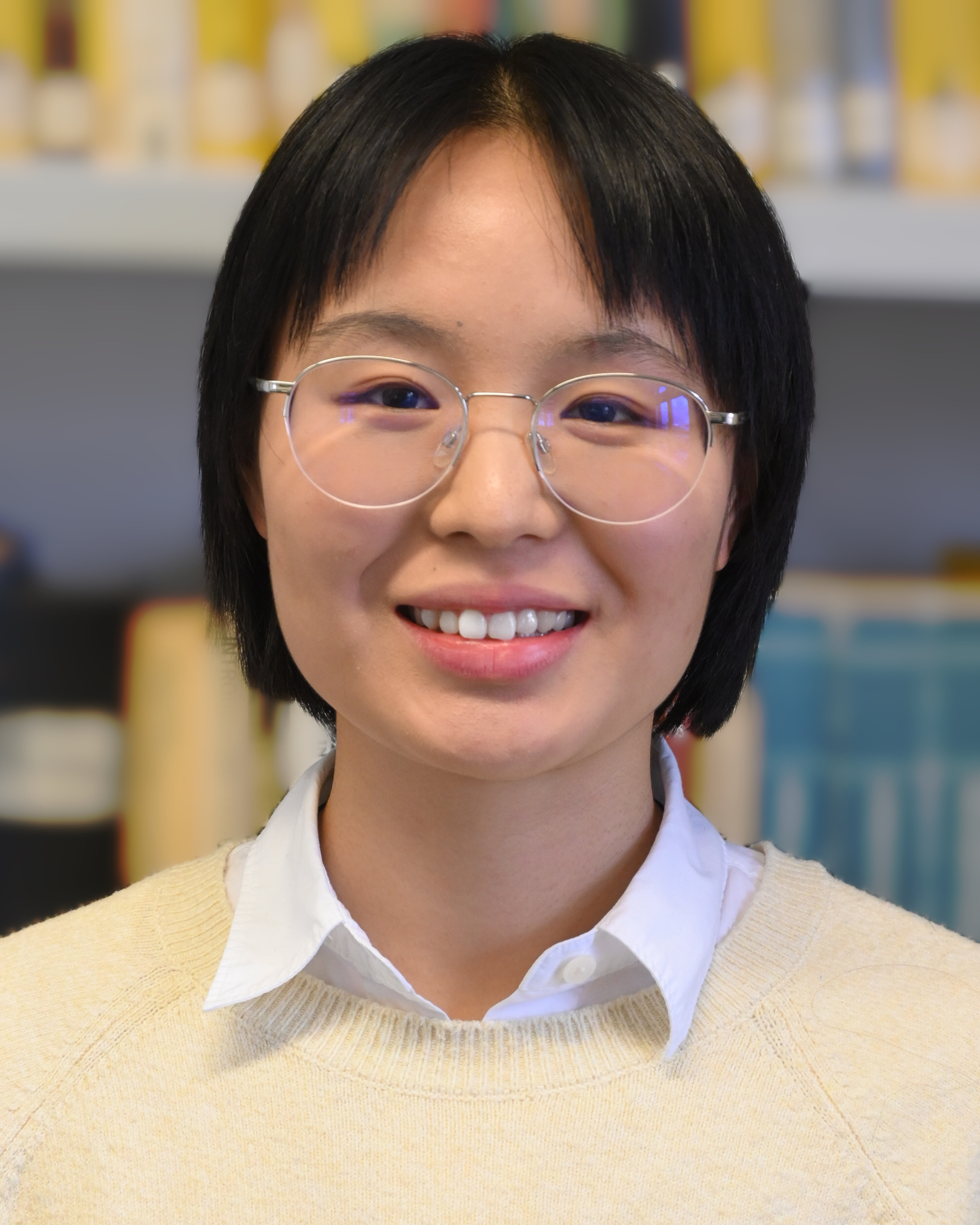}}]{Sisi~Miao} (S'21) completed her Bachelor's degree in Communication Engineering at Ocean University of China in 2018 and later obtained her Master's degree in INFOTECH from the University of Stuttgart in 2021. 
		
		She is currently a Ph.D. student at the Communications Engineering Laboratory, Karlsruhe Institute of Technology, focusing her research on classical and quantum error control coding.
	\end{IEEEbiography}
	\begin{IEEEbiography}[{\includegraphics[width=1in,height=1.25in,clip,keepaspectratio]{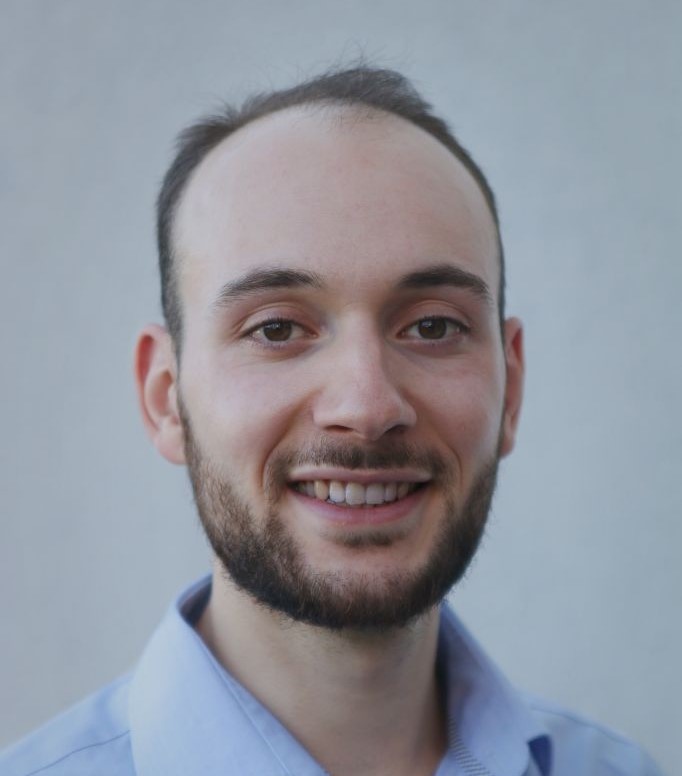}}]{Alexander~Schnerring} was born in Heidelberg, Germany in 1997. He received his B.Sc. and M.Sc. degree in electrical engineering from the Karlsruhe Institute of Technology in 2019 and 2022, respectively.
		
		He is currently working towards a PhD degree at the German Aerospace Center (DLR) Institute of Solar Research in Almería, Spain, where he is investigating the calibration of CST plants using autonomous UAVs. 
		His research interests include Signal Processing, Robotics and Computational Geometry. 
	\end{IEEEbiography}
	\begin{IEEEbiography}[{\includegraphics[width=1in,height=1.25in,clip,keepaspectratio]{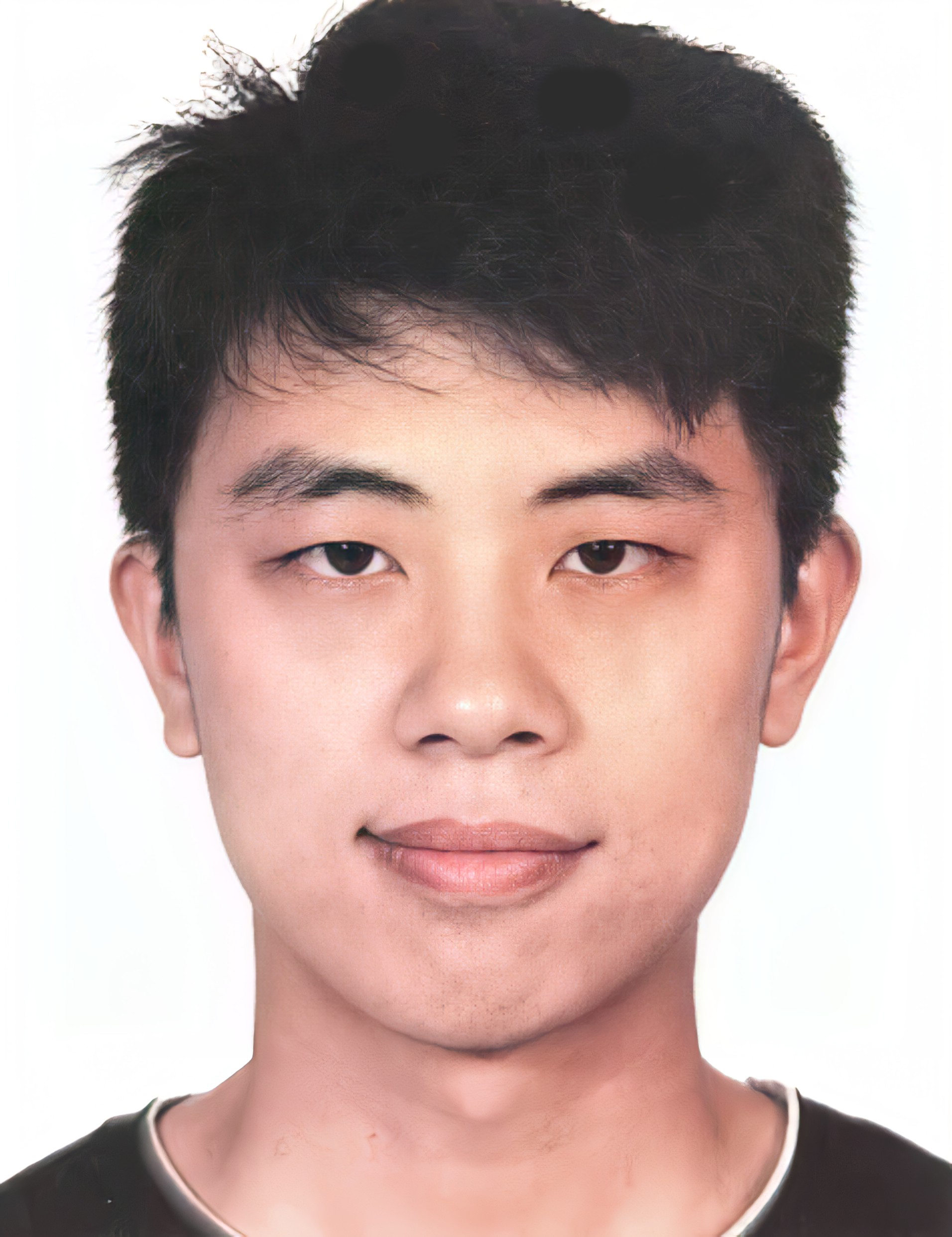}}]{Haizheng~Li} completed his Bachelor's degree in Automation at Beijing Institute of Technology in 2017 and later obtained his Master's degree in ETIT from the Karlsruhe Institute of Technology in 2021. 
		
		He is currently a Ph.D. student at the Communications Engineering Laboratory, Karlsruhe Institute of Technology, focusing his research on spatially-coupled LDPC codes.
	\end{IEEEbiography}
	\begin{IEEEbiography}[{\includegraphics[width=1in,height=1.25in,clip,keepaspectratio]{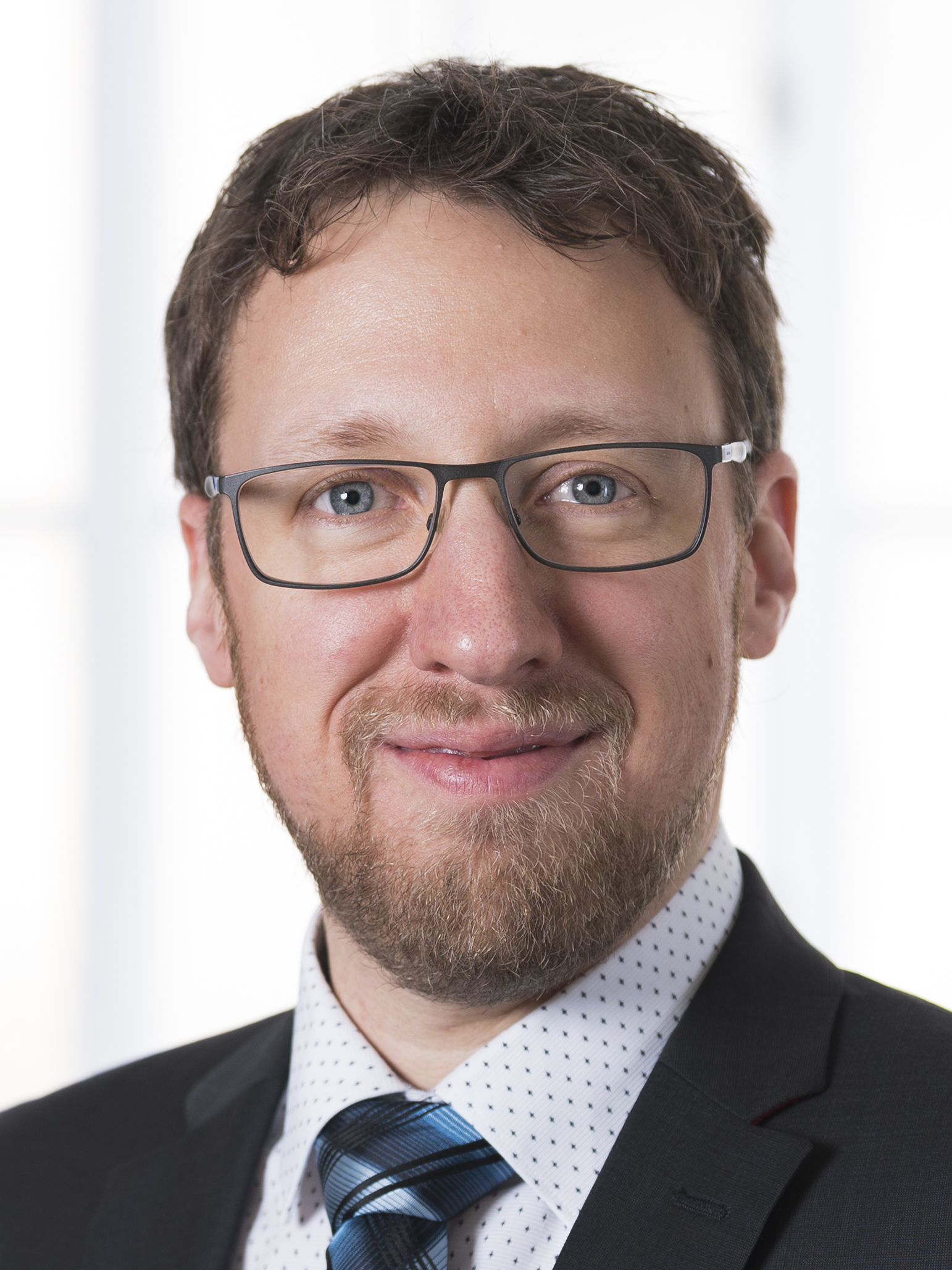}}]{Laurent~Schmalen} (S'06--M'13--SM'16--F'23)
		received the Diploma-(Ing.) in electrical engineering and information technology and the Dr.-(Ing.) degree from RWTH Aachen University, Germany. 
		
		From 2011 to
		2019, he was with Alcatel-Lucent Bell Labs and Nokia Bell Labs. From 2014 to 2019, he was also a Guest Lecturer with the University of Stuttgart, Stuttgart, Germany. Since 2019, he is Professor at
		the Karlsruhe Institute of Technology, where he co-heads the Communications
		Engineering Laboratory. His research interests include channel coding,
		modulation formats, and optical communications. 
		
		He was
		the recipient and co-recipient of several awards, including the E-Plus Award for
		his Ph.D. thesis, 2013 Best Student Paper Award from the IEEE Signal Processing
		Systems Workshop, and 2016 and 2018 Journal of Lightwave Technology
		best paper awards. He is an Associate Editor for the \textsc{IEEE Transactions on Communications}.
	\end{IEEEbiography}

	\EOD
	

\begin{thebibliography}{10}
		\bibitem{miao2023}
		S.~Miao, A.~Schnerring, H.~Li, and L.~Schmalen, ``Neural belief propagation decoding of quantum {LDPC} codes using overcomplete check matrices,'' in {\em Proc. IEEE \REV{Inf. Theory Workshop (ITW)}}, p.~TuA2.3, 2023.
		
		
		\bibitem{Gottesman14fault}
		D.~Gottesman, ``Fault-tolerant quantum computation with constant overhead,'' {\em Quantum Information and Computation}, vol.~14, pp.~1338--1372, 11 2014.
		
		\bibitem{tillich2013quantum}
		J.-P. Tillich and G.~Z{\'e}mor, ``Quantum LDPC codes with positive rate and minimum distance proportional to the square root of the blocklength,'' {\em IEEE Trans. Inf. Theory}, vol.~60, no.~2, pp.~1193--1202, 2013.
		
		\bibitem{panteleev2021quantum}
		P.~Panteleev and G.~Kalachev, ``Quantum {LDPC} codes with almost linear minimum distance,'' {\em IEEE Trans. Inf. Theory}, vol.~68, no.~1, pp.~213--229, 2021.
		
		\bibitem{panteleev2022asymptotically}
		P.~Panteleev and G.~Kalachev, ``Asymptotically good quantum and locally testable classical LDPC codes,'' in {\em Proc. Annual ACM SIGACT Symposium on Theory of Computing}, pp.~375--388, 2022.
		
		\bibitem{breuckmann21balanced}
		N.~P. Breuckmann and J.~N. Eberhardt, ``Balanced product quantum codes,'' {\em IEEE Trans. Inf. Theory}, vol.~67, no.~10, pp.~6653--6674, 2021.
		
		\bibitem{dennis2002topological}
		E.~Dennis, A.~Kitaev, A.~Landahl, and J.~Preskill, ``Topological quantum memory,'' {\em Journal of Mathematical Physics}, vol.~43, no.~9, pp.~4452--4505, 2002.
		
		\bibitem{edmonds1965paths}
		J.~Edmonds, ``Paths, trees, and flowers,'' {\em Canadian Journal of mathematics}, vol.~17, pp.~449--467, 1965.
		
		\bibitem{kuo2020refined}
		K.-Y. Kuo and C.-Y. Lai, ``Refined belief propagation decoding of sparse-graph quantum codes,'' {\em \REV{IEEE J. Sel. Areas Inf. Theory}}, vol.~1, no.~2, pp.~487--498, 2020.
		
		\bibitem{lai2021log}
		C.-Y. Lai and K.-Y. Kuo, ``Log-domain decoding of quantum LDPC codes over binary finite fields,'' {\em \REV{IEEE Trans. Quantum Eng.}}, vol.~2, pp.~1--15, 2021.
		
		\bibitem{panteleev2021degenerate}
		P.~Panteleev and G.~Kalachev, ``Degenerate quantum LDPC codes with good finite length performance,'' {\em Quantum}, vol.~5, p.~585, 2021.
		
		\bibitem{raveendran2021trapping}
		N.~Raveendran and B.~Vasi{\'c}, ``Trapping sets of quantum {LDPC} codes,'' {\em Quantum}, vol.~5, p.~562, 2021.
		
		\bibitem{rigby2019modified}
		A.~Rigby, J.~Olivier, and P.~Jarvis, ``Modified belief propagation decoders for quantum low-density parity-check codes,'' {\em Physical Review A}, vol.~100, no.~1, p.~012330, 2019.
		
		\bibitem{roffe2020decoding}
		J.~Roffe, D.~R. White, S.~Burton, and E.~Campbell, ``Decoding across the quantum low-density parity-check code landscape,'' {\em Physical Review Research}, vol.~2, no.~4, p.~043423, 2020.
		
		\bibitem{poulin2008iterative}
		D.~Poulin and Y.~Chung, ``On the iterative decoding of sparse quantum codes,'' {\em Quantum Information and Computation}, vol.~8, no.~10, pp.~987--1000, 2008.
		
		\bibitem{wang2012enhancedfeedback}
		Y.-J. Wang, B.~C. Sanders, B.-M. Bai, and X.-M. Wang, ``Enhanced feedback iterative decoding of sparse quantum codes,'' {\em IEEE Trans. Inf. Theory}, vol.~58, no.~2, pp.~1231--1241, 2012.
		
		\bibitem{crest2022stabilizer}
		V.~S. Julien Du~Crest, Mehdi~Mhalla, ``Stabilizer inactivation for message-passing decoding of quantum {LDPC} codes,'' in {\em Proc. IEEE \REV{Inf. Theory Workshop (ITW)}}, pp.~488--493, 2022.
		
		
		
		\bibitem{DM98}
		M.~Davey and D.~MacKay, ``Low density parity check codes over GF(q),'' in {\em Proc. Information Theory Workshop (ITW)}, pp.~70--71, 1998.
		
		\bibitem{DF07}
		D.~Declercq and M.~Fossorier, ``Decoding algorithms for nonbinary LDPC codes over GF$(q)$,'' {\em IEEE Trans. Commun.}, vol.~55, no.~4, pp.~633--643, 2007.
		
		\bibitem{liu2019neural}
		Y.-H. Liu and D.~Poulin, ``Neural belief-propagation decoders for quantum error-correcting codes,'' {\em Phys. Rev. Lett.}, vol.~122, no.~20, p.~200501, 2019.
		
		\bibitem{gottesman1997stabilizer}
		D.~Gottesman, {\em Stabilizer codes and quantum error correction}.
		\newblock PhD thesis, California Institute of Technology, 1997.
		
		\bibitem{Calderbank1998quantum}
		A.~R. Calderbank, E.~M. Rains, P.~W. Shor, and N.~J.~A. Sloane, ``Quantum error correction via codes over {GF}(4),'' {\em IEEE Trans. Inf. Theory}, vol.~44, no.~4, pp.~1369--1387, 1998.
		
		\bibitem{kitaev2006anyons}
		A.~Kitaev, ``Anyons in an exactly solved model and beyond,'' {\em Ann. Phys.}, vol.~321, no.~1, pp.~2--111, 2006.
		
		\bibitem{manabu2012}
		M.~Hagiwara, M.~P.~C. Fossorier, and H.~Imai, ``Fixed initialization decoding of {LDPC} codes over a binary symmetric channel,'' {\em IEEE Trans. Inf. Theory}, vol.~58, no.~4, pp.~2321--2329, 2012.
		
		\bibitem{Leon88}
		J.~S. Leon, ``A probabilistic algorithm for computing minimum weights of large error-correcting codes,'' {\em IEEE Trans. Inf. Theory}, vol.~34, no.~5, pp.~1354--1359, 1988.
		
		\bibitem{lian2019learned}
		M.~Lian, F.~Carpi, C.~H{\"a}ger, and H.~D. Pfister, ``Learned belief-propagation decoding with simple scaling and SNR adaptation,'' in {\em Proc. IEEE Int. Symp. Inf. Theory (ISIT)}, pp.~161--165, 2019.
		
		\bibitem{halford2006random}
		T.~R. Halford and K.~M. Chugg, ``Random redundant soft-in soft-out decoding of linear block codes,'' in {\em IEEE Int. Symp. Inf. Theory (ISIT)}, pp.~2230--2234, 2006.
		
		\bibitem{bossert1986hard}
		M.~Bossert and F.~Hergert, ``Hard-and soft-decision decoding beyond the half minimum distance---an algorithm for linear codes (corresp.),'' {\em IEEE Trans. Inf. Theory}, vol.~32, no.~5, pp.~709--714, 1986.
		
		\bibitem{kothiyal2005iterative}
		A.~Kothiyal, O.~Y. Takeshita, W.~Jin, and M.~Fossorier, ``Iterative reliability-based decoding of linear block codes with adaptive belief propagation,'' {\em IEEE Commun. Lett.}, vol.~9, no.~12, pp.~1067--1069, 2005.
		
		\bibitem{Jiang2006iterative}
		J.~Jiang and K.~R. Narayanan, ``Iterative soft-input soft-output decoding of {R}eed--{S}olomon codes by adapting the parity-check matrix,'' {\em IEEE Trans. Inf. Theory}, vol.~52, no.~8, pp.~3746--3756, 2006.
		
		\bibitem{buchberger2020pruning}
		A.~Buchberger, C.~H{\"a}ger, H.~D. Pfister, L.~Schmalen, and A.~Graell~i~Amat, ``Pruning and quantizing neural belief propagation decoders,'' {\em IEEE J. Sel. Areas Commun.}, vol.~39, no.~7, pp.~1957--1966, 2020.
		
		\bibitem{nachmani2016learning}
		E.~Nachmani, Y.~Be'ery, and D.~Burshtein, ``Learning to decode linear codes using deep learning,'' in {\em Proc. Annual Allerton Conference on Communication, Control, and Computing (Allerton)}, pp.~341--346, 2016.
		
		\bibitem{nachmani2018deep}
		E.~Nachmani, E.~Marciano, L.~Lugosch, W.~J. Gross, D.~Burshtein, and Y.~Be’ery, ``Deep learning methods for improved decoding of linear codes,'' {\em IEEE J. Sel. Top. Signal Process.}, vol.~12, no.~1, pp.~119--131, 2018.
		
		\bibitem{xiao2019neural}
		X.~Xiao, ``Neural-net decoding of quantum {LDPC} codes with straight-through estimators,'' in {\em Proc. Information Theory and Applications Workshop (ITA)}, 2019.
		
		\bibitem{goodfellow2016deep}
		I.~Goodfellow, Y.~Bengio, and A.~Courville, {\em Deep Learning}.
		\newblock MIT press, 2016.
		
		\bibitem{mackay2004sparse}
		D.~J. MacKay, G.~Mitchison, and P.~L. McFadden, ``Sparse-graph codes for quantum error correction,'' {\em IEEE Trans. Inf. Theory}, vol.~50, no.~10, pp.~2315--2330, 2004.
		
		\bibitem{higgott2023sparse}
		O.~Higgott and C.~Gidney, ``Sparse blossom: correcting a million errors per core second with minimum-weight matching,'' {\em arXiv preprint arXiv:2303.15933}, 2023.
		
		\bibitem{8006751}
		L.~Lugosch and W.~J. Gross, ``Neural offset min-sum decoding,'' in {\em Proc. IEEE Int. Symp. Inf. Theory (ISIT)}, pp.~1361--1365, 2017.
		
		
		\bibitem{ref1}Q. Lu, Z. Shen, C. -W. Sham and F. C. M. Lau, "A parallel-routing network for reliability inferences of single-parity-check decoder," in {\em Proc. Int. Conf. on Advanced Technologies for Communications (ATC)}, pp. 127--132, 2015.
		
		\bibitem{ref2}C. -W. Sham, X. Chen, W. M. Tam, Y. Zhao and F. C. M. Lau, "A layered QC-LDPC decoder architecture for high speed communication system," in {\em Proc. IEEE Asia Pacific Conf. on Circuits and Systems}, pp. 475-478, 2012.
		
		
	\end{thebibliography}
\end{document}